\newcommand{\PaperNumber}{XXX}
\documentclass[10pt,sigconf,letterpaper,nonacm]{acmart}

\usepackage{subfig} 
\usepackage{multirow} 
\usepackage{pifont} 
\usepackage{fontawesome} 
\usepackage[ruled,linesnumbered]{algorithm2e} 
\usepackage[none]{hyphenat}

\begin{document}

\title{ZBanner: Fast Stateless Scanning Capable of Obtaining Responses over TCP}


\author{Chiyu Chen}
\affiliation{%
	\orcid{0009-0009-5435-3083}
	\institution{College of Electronic Engineering, National University of Defense Technology}
	\city{Hefei}
	\state{Anhui}
	\country{China}}
\email{chenchiyu14@nudt.edu.cn}

\author{Yuliang Lu}
\affiliation{%
	\institution{College of Electronic Engineering, National University of Defense Technology}
	\city{Hefei}
	\state{Anhui}
	\country{China}}
\email{luyuliang@nudt.edu.cn}

\author{Guozheng Yang}
\affiliation{%
	\institution{College of Electronic Engineering, National University of Defense Technology}
	\city{Hefei}
	\state{Anhui}
	\country{China}}
\email{yangguozheng17@nudt.edu.cn}

\author{Yi Xie}
\affiliation{%
	\institution{College of Electronic Engineering, National University of Defense Technology}
	\city{Hefei}
	\state{Anhui}
	\country{China}}
\email{xie\_yi@nudt.edu.cn}

\author{Shasha Guo}
\affiliation{%
	\institution{College of Electronic Engineering, National University of Defense Technology}
	\city{Hefei}
	\state{Anhui}
	\country{China}}
\email{guoshasha13@nudt.edu.cn}


\begin{abstract}

\sloppy{}

Fast large-scale network scanning is an important way to understand internet service configurations and security in real time, among which stateless scan is representative.
Existing stateless scanners can perform single-packet scans for internet-wide network measurements but are limited to host discovery or port scanning. To obtain further information over TCP, slower stateful scanners must be used in conjunction which spend more time and memory because of connection state maintenance.
Through simplifying TCP finite state machine, this paper proposes a novel stateless scanning model, which can establish TCP connections and obtain further responses in a completely stateless manner.
Based on this model, we implement ZBanner, an improved modular stateless scanner that utilizes user-defined probes for identifying services and versions, fingerprinting TLS servers, etc.
We present unique design of ZBanner and experimentally characterize its feasibility and performance.
Experiments show that ZBanner performs better than current state-of-the-art solutions in terms of scan rate and memory usage.
ZBanner achieves at least three times faster than current tools for generic ports and over 90 times faster for open ports while keeping a minimum and stable memory usage.

\end{abstract}


\maketitle

\section{Introduction}

\sloppy{}

Fast large-scale network scanning serves as an important way for understanding service configurations and security within the internet in real time. Among them, stateless scan is often used to quickly collect information of internet-scale ports, which is the basis of subsequent network measurements\cite{diffieHellman, missionHTTP, drown, icsview, ftpCloud, effectiveVul}.
Recent research indicates that focusing just on TCP liveness during large-scale network surveys, which is what stateless scanners do, can lead to misjudgments on actual services\cite{lzr} and overlook that some ports cannot complete the expected application-layer handshakes. To address these limitations, many studies incorporate stateful scanners which are capable of establishing complete TCP connections to obtain more accurate and comprehensive information about target ports\cite{pumpjarm, llmfingerprint, firmfinger2, chargeprint, cardiscovery, vulanalysis, routerVendor}.

Existing stateless scanners are limited to single-packet scan such as host discovery or port scanning without local state maintaining. They cannot obtain further information over TCP.
Stateful scanners, which can establish TCP connections with target ports to obtain further responses, have to maintain states for every connections through protocol stack. 
However, when conducting large-scale scans, maintaining a significant number of connection states becomes resource-intensive and also slows down the scan rate especially on ordinary hardware configuration.
Consequently, two-phase scanning that combines stateless and stateful scanners are typically used and spend more time and resources overall.

In this paper, we first analyze popular stateless scanning techniques from the perspective of TCP finite state machine (FSM). Then, we simplify the FSM for TCP communication and construct a stateless scanning model, which can do TCP connection and recieve further responses in completely stateless manner. Based on this model, we implement ZBanner—a modular stateless scanner designed for fast large-scale network measurement. ZBanner establishes connections with target ports and achieves high-speed responses obtaining over TCP with low memory usage. Specifically, ZBanner remains robust from a number of network anomalies during the scanning process. In addition, ZBanner includes extensible probe modules which enables various scanning tasks.

Through a series of experiments and comparisons with state-of-the-art solutions, we characterized ZBanner's feasibility and validated its performance on scan rate and memory usage for obtaining responses over TCP in a most generalized scenarios. Experimental results demonstrate that ZBanner achieves 98\% coverage, which is independent of the scan rate and sufficiently comprehensive for typical research applications. 
Furthermore, ZBanner's scan rate and memory usage advantages increase as the proportion of known open ports rises. On odinary hardware configuration, ZBanner performs at least three times faster than current state-of-the-art solutions. When scanning specific target ports already determined to be open, ZBanner reduces scan time by over 90 times compared to existing approaches, all while maintaining minimal and stable memory usage.

This paper makes the following key contributions:
\begin{itemize}
 	\item We construct a stateless scanning model through simplifying TCP FSM, which is the first work that is able to establish TCP connection and obtain further responses in a completely stateless manner.
 	\item Based on the stateless scanning model, we implement a modular stateless scanner called ZBanner for fast large-scale network measurement. Depended on user-defined probes, ZBanner is capable of banner grabbing, service identification or version detection efficiently, etc.
 	\item Through a series of experiments and comparisons, we validate the feasibility of stateless scanning model and demonstrate that ZBanner outperforms current state-of-the-art solutions.
\end{itemize}

\section{Related Works}

\sloppy{}

Our work improves on existing stateless scanning technique to enable stateless scanner for establishing TCP connections and obtaining further responses. This makes stateless scanner operates just like two-phase scan but with a faster scan rate and lower memory usage.
In this section, we introduce existing well-known stateless scanning technique and commonly used two-phase scan solutions. The latter will be the subject of comparison in our experimental part.

\subsection{Stateless Scanning Technique}

Stateless scan bypasses the TCP protocol stack of operating system by sending ethernet layer packets as probes via a raw socket.
Take the TCP SYN scan as an example, these probes utilize an encoding scheme similar to SYN cookies for response packets matching. 
By eliminating local state maintenance, this technique avoids memory limitations and complex state tranform process so that accelerates scanning speeds across the internet.
Stateless scanning technique serves as the foundation for large-scale network measurements and is the initial step in identifying vulnerable nodes or weakness on the internet\cite{spoki}. ZMap\cite{zmap}, the most popular stateless port scanner, efficiently utilizes network interfaces and bandwidth resources to scan specified ports across the entire IPv4 address space within 45 minutes. Similar to ZMap, XMap\cite{xmap} improved by Li et al. and open-source tools like Masscan\cite{masscan} also employ similar stateless scanning technique with additional support for IPv6. Building upon these tools, Yarrp\cite{yarrp2016,yarrp2018}, Flashroute\cite{flashroute}, and D-Miner\cite{dminer} apply Stateless scanning technique to network topology discovery, achieving rapid tracking of \verb|/24| prefix routes on an internet scale.

However, existing stateless scanners are designed for single-packet scan and suitable only for limited scenarios such as host discovery and port scanning that work without complete TCP connections.

\subsection{Two-Phase Scan}

In real-world scenarios, stateless port scanners are often combined with slower stateful scanners which are capable of establishing complete TCP connections\cite{spoki} to obtain further information. These two-phase approaches decompose the scanning process into port scanning and further information obtaining over TCP.

Some studies optimize scanning mechanisms to achieve rapid scan rate for stateful scanners as much as possible. Nmap\cite{nmap}, for instance, employs its custom asynchronous communication library \verb|Nsock| to obtain information over TCP from target ports, forming the basis for features like service identification and version detection. ZGrab\cite{zgrab} utilizes Go language’s standard library \verb|Net| and goroutines to expect a high-speed scan. Actually, ZGrab relies on OS provided asynchronous communication API. LZR\cite{lzr} and Masscan, aiming for service identification, maintain TCP connection states in user space memory to complete scanning after confirming port availability without new connection.

Depending on whether the stateless and stateful scanners share a single TCP connection for one target port, two-phase scanning approach can be further categorized into \textbf{continuous two-phase scan} and \textbf{separate two-phase scan}. ZMap/LZR and ZMap/ZGrab are representive solutions for these two scanning types and widely used in various large-scale network measurements.

However, the introduction of stateful scanners in two-phase scan leads to dependence on complete TCP connections through operating system or user-space (maybe incomplete) TCP protocol stack. Although many studies leverage asynchronous mechanisms to enhance its efficiency, maintaining a large number of TCP connection states during large-scale scanning limits overall scan rate and spends more memory .


\section{Stateless Scanning Model}

\sloppy{}

Existing stateless scanning technique only support single-packet scan and collect limited information. Our target is contructing a new stateless scanning model which can establish complete TCP connections for further information obtaining.
In this section, we analyze the principle of existing stateless scanning technique implemented by popular port scanner ZMap from a perspective of TCP finite state machine (FSM). Drawing from this analysis, we simplify the TCP FSM for typical scanning scenarios involving TCP communication, ultimately constructing a stateless scanning model which are capable of completing TCP connections and obtaining further responses.

\subsection{Analysis to Existing Stateless Machanism}\label{sec:analyzezmap}

ZMap efficiently utilizes network interfaces and bandwidth due to three key factors:

\begin{itemize}
	\item ZMap encodes state information into SYN packets, achieving statelessness during scanning.
	\item ZMap separates packet send and receive operations into two threads, minimizing propagation delays.
	\item ZMap creates a permutation covering the entire target address space, effectively distributing scans for every subnet.
\end{itemize}

The permutation algorithm mitigates pressure on target subnets. Additionally, asynchronous communication mechanisms are not unique to ZMap; mainstream operating systems already provide asynchronous communication interfaces (e.g., \verb|select|, \verb|poll|, and \verb|epoll|). Therefore, ZMap’s key optimization lies in achieving statelessness during the scanning process compared to other scanners.

\begin{figure}[htbp]
	\centering
	\subfloat[Communication Process of ZMap]
	{
		\includegraphics[width=0.4\columnwidth]{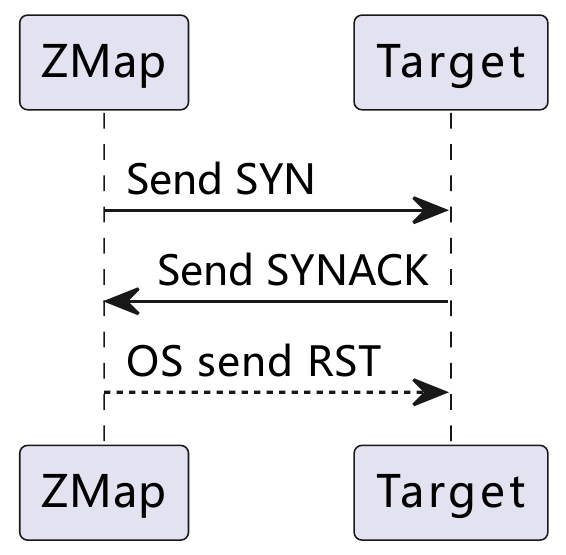}
		\label{fig:zmapcomm}
	}
	\subfloat[FSM of ZMap]
	{
		\includegraphics[width=0.4\columnwidth]{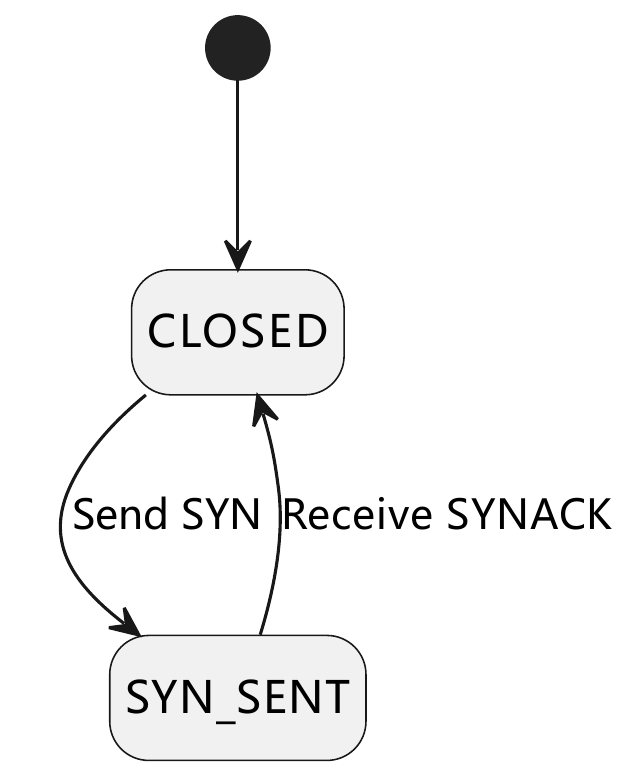}
		\label{fig:zmapfsm}
	}
	\caption{Analysis to Stateless Mechanism of ZMap}
	\label{fig:zmapfromfsm} 
\end{figure}

ZMap utilizes TCP SYN scan to detect open ports which is also known as half-open port scan involving sending TCP SYN packets to target ports and determining their openness based on the response type. Without full TCP connections that involve further data exchange, TCP SYN scan remains at the TCP handshake stage from an FSM perspective. Additionally, ZMap bypasses the operating system’s TCP/IP protocol stack, establishing a half-connection with the target port and prompting an automatic RST response from the Linux kernel. Consequently, ZMap eliminates the need for an additional packet transmission during scanning, as illustrated in the Figure \ref{fig:zmapcomm}.

By simplifying the communication process for specific scenarios, ZMap operates with only two states during scanning: \verb|CLOSED| and \verb|SYN_SENT|. The state transition diagram in the Figure \ref{fig:zmapfsm} reflects this simplicity. With just two states, ZMap identifies the \verb|SYN_SENT| state by encoding information into the sequence number field of SYN packets. This approach leverages the deterministic relationship between the acknowledgment numbers in received SYN-ACK or RST packets and the \verb|SYN_SENT| state.

\subsection{Construction of Scanning Model}

To obtain further information over TCP, scanning process must involve complete TCP connection establishing with target ports, specified probes sending, responses receiving, and connection closing. Although existing work has employed target permutation algorithms and asynchronous communication mechanisms, the complexity of scanning process to obtain information over TCP necessitates maintaining connection states in either the kernel or user space. Unfortunately, the process of state maintenance is dispersed into intervals between packet sending and receiving. This consumes substantial runtime memory and restricts the maximum scan rate. Achieving complete statelessness in this scenario would significantly enhance scanning efficiency.

Inspired by the analysis in Section \ref{sec:analyzezmap}, we attempt to simplify the TCP FSM based on scanning scenarios that obtaining information over TCP. Our simplification process adheres to the following principles as shown in Figure \ref{fig:fsmSimplify}:

\begin{figure*}[htbp]
	\centering
	\includegraphics[width=1.5\columnwidth]{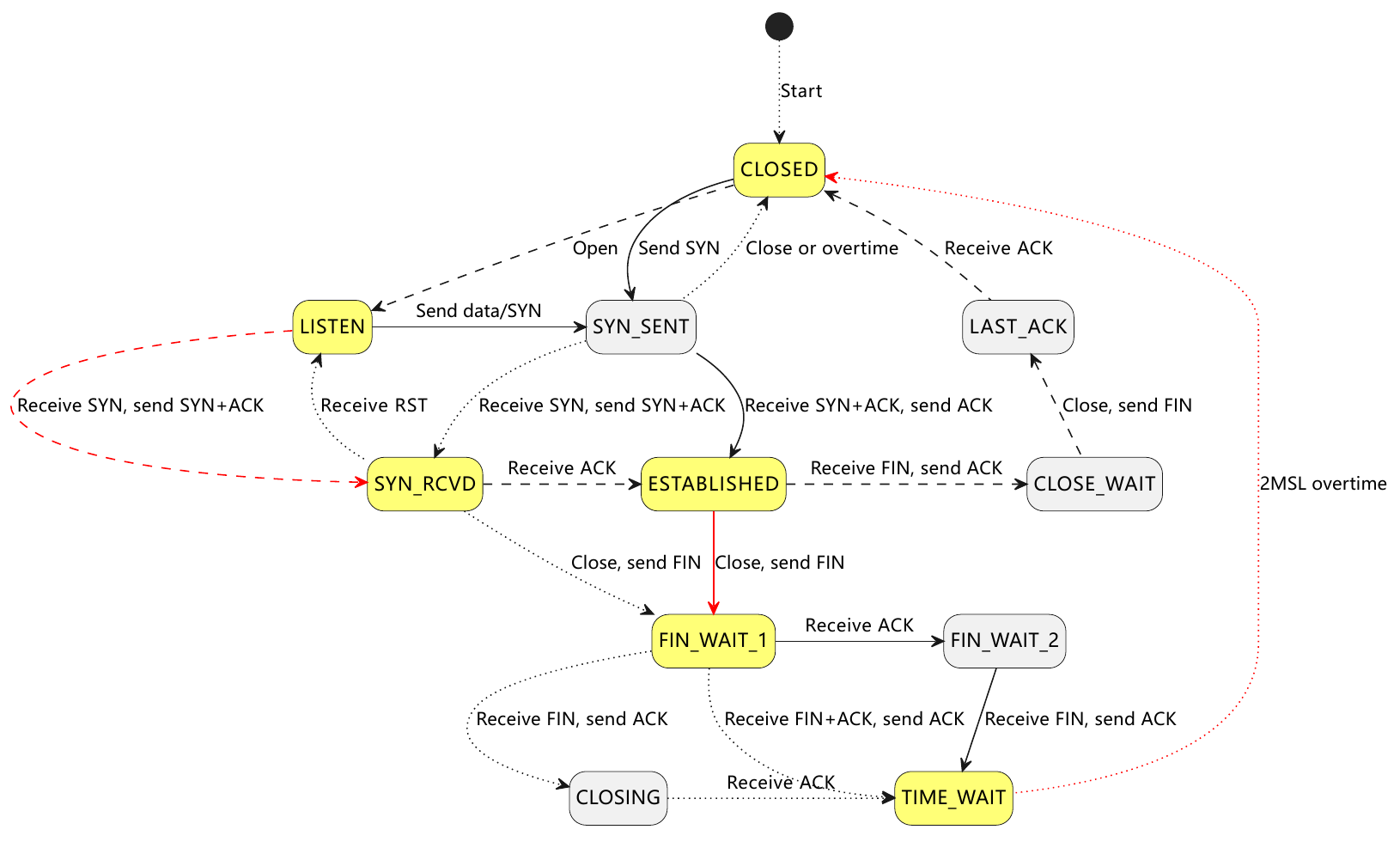}
	\caption{Simplification of FSM in Specific Scanning Scenarios}
	\label{fig:fsmSimplify}
\end{figure*}

\begin{enumerate}
	\item Since scanning is a client-side behavior, we remove typical server-side state transitions, such as from \verb|LISTEN| to \verb|SYN_RCVD|.
	\item Stateless conditions prevent us from maintaining timers for connections. Consequently, we eliminate state transitions caused by timeouts, such as from \verb|TIME_WAIT| to \verb|CLOSED|.
	\item In the absence of maintaining half-closed connections, we always choose to terminate communication by having the client send an RST packet. This simplifies the process such as from \verb|ESTABLISHED| to \verb|FIN_WAIT_1|.
\end{enumerate}

Following the outlined approach, we arrive at a preliminarily simplified FSM as depicted in the Figure \ref{fig:prefinalfsm} and the details of simplifying FSM can be found in Appendix \ref{sec:fsmSimpify}. Unlike existing stateless scanning technique, ours involve establishing complete TCP connection states and subsequent interactions. Consequently, we introduce the \verb|ESTABLISHED| state\cite{establishfsm} into the FSM.

\begin{figure}[htbp]
	\centering
	\includegraphics[width=1.0\columnwidth]{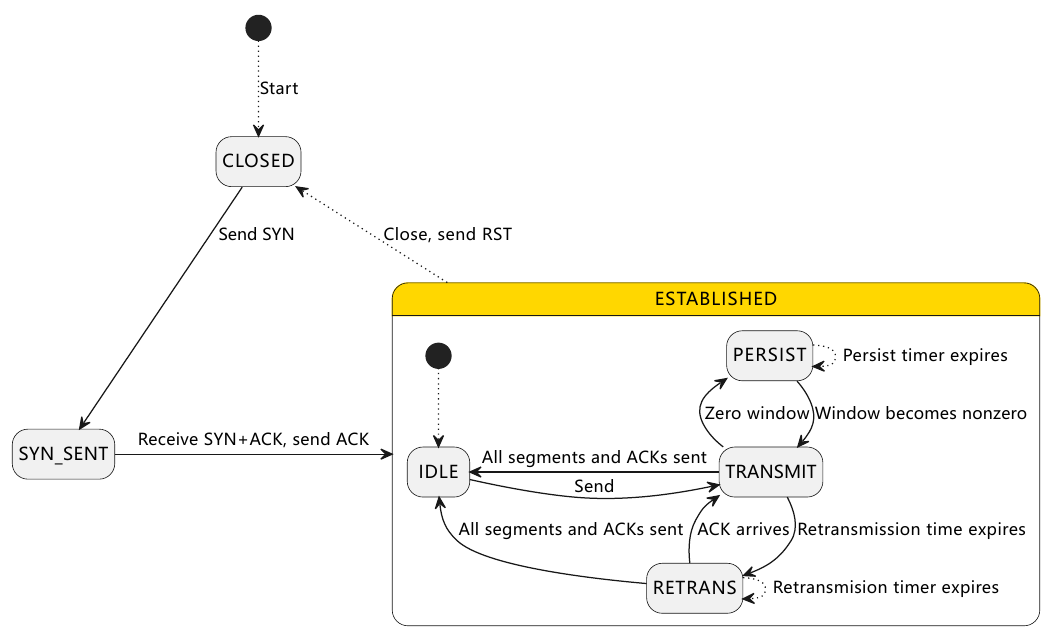}
	\caption{Preliminarily Simplified FSM}
	\label{fig:prefinalfsm}
\end{figure}

\begin{figure}[htbp]
	\centering
	\subfloat[Typical Pattern of Communication Process]
	{
		\includegraphics[width=0.4\columnwidth]{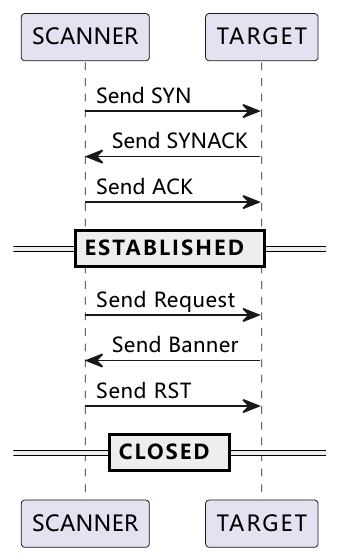}
		\label{fig:connsimplify1}
	}
	\subfloat[Merging the Handshake and ESTABLISHED Phases]
	{
		\includegraphics[width=0.45\columnwidth]{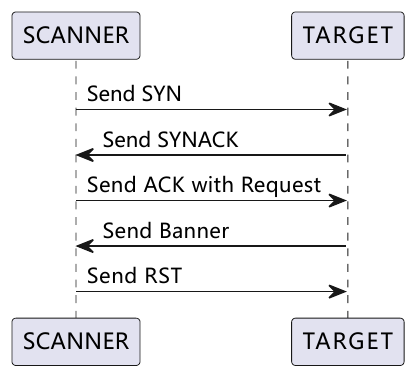}
		\label{fig:connsimplify2}
	}
	\caption{Simplification of Communication Process in Scanning}
	\label{fig:connsimplify} 
\end{figure}

\begin{figure}[htbp]
	\centering
	\includegraphics[width=1.0\columnwidth]{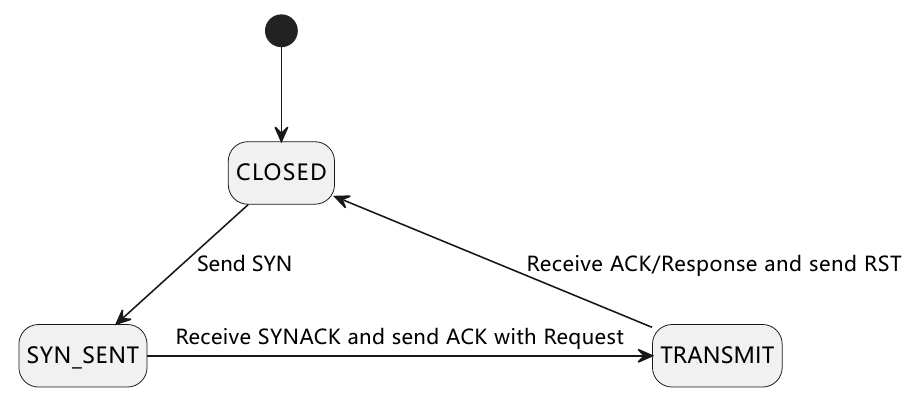}
	\caption{Simplified FSM}
	\label{fig:finalfms}
\end{figure}

Although the internal transitions within \verb|ESTABLISHED| are complex, we do not require long-term stable TCP data communication while scanning. Thus, we avoid maintaining congestion control and retransmission mechanisms to simplify the general communication pattern of scanning (as shown in the Figure \ref{fig:connsimplify1}). By utilizing the Piggybacking\cite{computernetworking} to attach probe payloads to ACK packets during the three-way handshake, we further streamline the communication process (as illustrated in Figure \ref{fig:connsimplify2}). At this point, the handshake and \verb|ESTABLISHED| stages are no longer explicitly differentiated, resulting in the final FSM as shown in the Figure \ref{fig:finalfms}.

\begin{figure}[htbp]
	\centering
	\includegraphics[width=1.0\columnwidth]{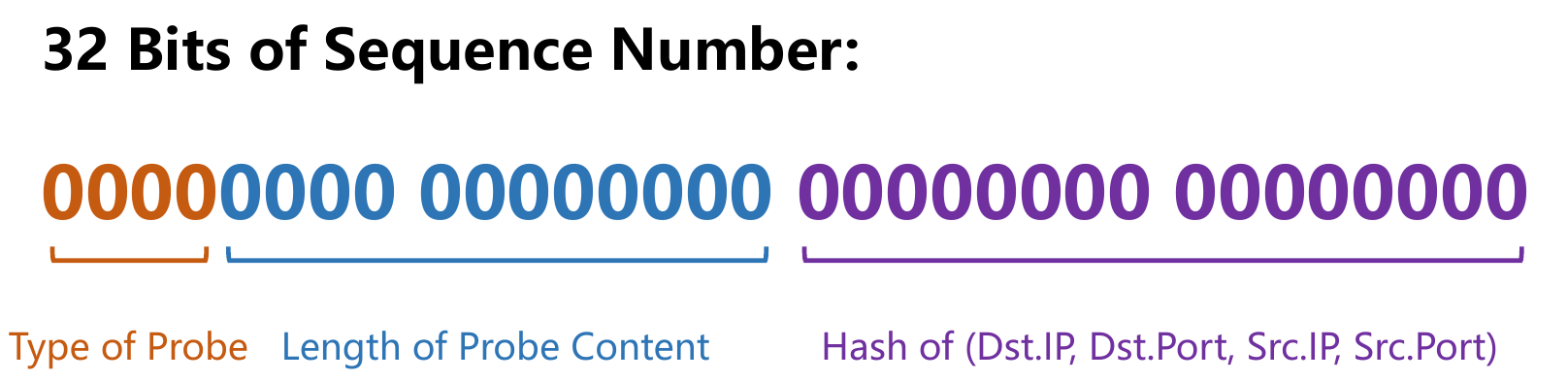}
	\caption{Encoding of Sequence Number}
	\label{fig:cookiecode}
\end{figure}

The simplified FSM for scanning comprises three states: \verb|CLOSED|, \verb|SYN_SENT|, and \verb|TRANSMIT|. While \verb|SYN_SENT| is identified using the relationship between the SYN packet’s sequence number and the received SYN-ACK packet’s acknowledgment number, \verb|TRANSMIT| is identified based on the difference between their acknowledgment numbers (which corresponds to the length of probe content). To support multiple probe types, we encode the probe type, content length, and TCP connection quadruple hash into the initial sequence number of the SYN packet, as depicted in Figure \ref{fig:cookiecode}.

After identifying all states, we construct the final stateless scanning model. The core algorithm (logic for the receiver thread) is illustrated in Algorithm \ref{algo:appfsm}. When we receive a TCP packet with ACK flag, we identify and handle its state as follows:

\begin{enumerate}
	\item If the acknowledgment number’s hash part (last 16 bits) equals the hash value of the TCP connection quadruple plus 1, we identify the packet as originating from a \verb|SYN_SENT| connection. We promptly send an ACK packet with the probe payload attached.
	\item When the acknowledgment number’s hash part equals the hash value of the TCP connection quadruple plus 1 and the acknowledgment number’s probe content length (bits 4 to 11), we identify the packet as coming from an \verb|ESTABLISHED| connection. We then determine the probe type based on the first 4 bits of the acknowledgment number. The accompanying data is processed by the corresponding probe module, and we actively send an RST packet to terminate the connection.
	\item For other acknowledgment numbers, the packet is either not a response from our initiated connection or indicates an error.
\end{enumerate}

\begin{algorithm}[htbp]
	\caption{Stateless Scanning Model}
	\label{algo:appfsm}
	\SetAlgoLined
	\KwIn{packet, probe, src, dst}
	
	\While{receive packet with ack}{
		hash = Hash(src.ip, src.port, dst.ip, src.port)\;
		probe\_type = packet.seqno\textgreater\textgreater 28\;
		content\_len = packet.seqno\textgreater\textgreater 16 \& 0xFFF\;
		\uIf{packet.ackno \& 0xFFFF == hash + 1 \&\& packet.flags.syn}{
			\tcp{Current state is SYN\_SENT}
			send\_ack\_with\_payload(probe\_type)\;
		}
		\uElseIf{packet.ackno \& 0xFFFF == hash + content\_len + 1}
		{
			\tcp{Current state is TRANSMIT}
			banner = read\_payload(packet)\;
			handle\_banner(probe\_type, banner)\;
			send\_rst()\;
		}
		\Else{
			\tcp{Packets we don't care}
			handle\_other\_packets\;
		}
	}
\end{algorithm}

\section{Design of ZBanner}

\sloppy{}

\begin{figure*}[htbp]
	\centering
	\includegraphics[width=1.8\columnwidth]{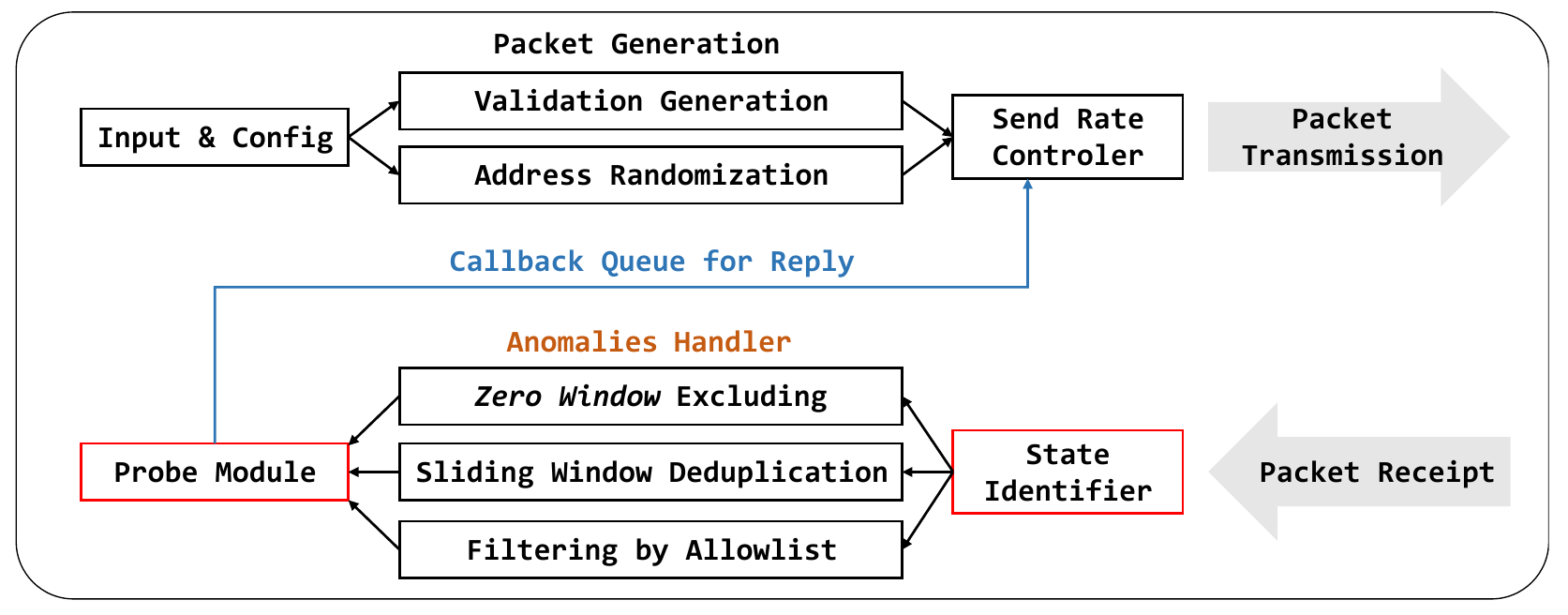}
	\caption{Architecture of ZBanner}
	\label{fig:zbanner}
\end{figure*}

Based on the stateless scanning model proposed in previous section, we designed and implemented a modular stateless scanner called ZBanner for large-scale network measurement. ZBanner is capable of doing port scanning and obtaining further information over TCP efficiently in a completely stateless manner with an architecture shown in Figure \ref{fig:zbanner}. Different from existing scanners:

\begin{itemize}
	\item ZBanner's state identifier can identify the two states mentioned in our stateless scanning model and handle other conditions.
	\item ZBanner's sender and receiver threads are not completely independent, but are associated via a callback queue to implement our stateless scanning model.
	\item ZBanner handles multiple connection anomalies in several ways while scanning.
	\item ZBanner uses extensible probe modules to achieve various tasks.
\end{itemize}

In this section, we describe the main unique designs of ZBanner for implementing the stateless scanning model.

\subsection{Asynchronous Mode with Callback Queue}

Current asynchronous stateless scanners decouple the packets sending and receiving into two separate threads. Leveraging the automatic RST reply feature of Linux, these tools allow the sender and receiver threads operating independently. However, in the context of our stateless scanning model, further packets need to be sent/reply to the target host after identified different states. Consequently, existing asynchronous modes do not meet our requirement.

To adapt ZBanner to our stateless scanning model, we introduce a callback queue between the sender and receiver threads based on the existing asynchronous mode. After the receiver thread received a packet and identified the state, it generates new packets that need to be sent into the callback queue. The sender thread will periodically take the packets from callback queue and send them actually.

To prevent connection anomalies due to timeouts, the sender thread prioritizes transmiting packets from the callback queue in each batch. When the callback queue becomes empty, the sender thread continues transmiting TCP SYN packets for new target ports. By utilizing an asynchronous mode with a callback queue, ZBanner is capable of replying appropriately for different states, thus achieving the stateless operation.

\subsection{Handling Various Connection Anomalies}

While our stateless scanning model is an idealized creation based on normal communication, ZBanner must handle various connection anomalies efficiently and robustly from the real world. Recent research indicates that across 37 popular ports, up to 96\% of services (per port) fail to complete the expected application-layer handshakes. Furthermore, even after establishing a TCP connection successfully like ZBanner, anamolies over TCP connection still exists during scanning. 

Combining several anomalies described by Izhikevich et al.\cite{lzr} and our findings in actual scans, Scanners may encounter the following connection anomalies that result in slowing down the scan rate or missing some targets.

\begin{enumerate}
	\item Connection Shunning: During the execution of a seperate two-phase scan, ZMap may cause a semi-connected state to target host, resulting in the failure of continuous stateful scan. This makes two-phase scan missing some target ports.
	\item Zero Window DDoS Protections: Target hosts may specify a zero window size in the TCP SYN-ACK packet to prevent us from sending further probes. This makes existing stateless scanners to believe the port is open and stateful scanners to keep waiting senselessly for the target to open its TCP window.
	\item Dropping Connections Mid-Handshake: Selective discarding of packets by middleboxes can lead to the target hosts not received the final ACK packets during TCP handshake process and prevent us from establishing connections. This makes stateful scanners receiving several TCP SYN-ACK packets because of the retransmiting of target and keep trying to reply TCP ACK packets.
	\item Dynamic Blocking after Handshake: Due to blocking of ACK packets from target port by middleboxes, target hosts cannot acknowledge the probe data we sent. This makes stateful scanners keep retransmitting probe senselessly.
	\item Waiting Complete Probe: When our probe contains data that does not match the service of target port yet, the target hosts confirms our probe and expects further data instead of responding an error to hint. This makes stateful scanners maintaining the connection for a long time until target host close it.
\end{enumerate}

Since ZBanner implements our stateless scanning model, it is inherently capable of handling some connection anomalies. For the "connection shunning", ZBanner continues to reuse the same connection to send probe and receive response over TCP after port scanning. It establishes only connection once per target port, thereby avoiding the target host entering a half-connected state and preventing continuous interaction failures. For the "dynamic blocking after handshake", stateless scanning model ensures that just valid connection state is maintained.

\textbf{Zero Window Exclusion}.
Regarding "zero window DDOS protection", if ZBanner receives a TCP SYN-ACK packet with a window size of zero during port scanning, it identifies this and actively closes the invalid connection.

\textbf{Sliding Window Deduplication}.
In scenarios of "dropping connections mid-handshake", where the target host does not receive the final ACK packet during the TCP handshake phase, SYN-ACK packets are retransmitted for multiple times. Within the receiver thread, ZBanner employs a sliding window mechanism to filter out duplicate packets within a short time frame during both port scanning and response obtaining over TCP. This approach prevents sending useless duplicate probes and conserve bandwidth. 

\textbf{Allowlist Filtering}.
For "waiting complete probe", ZBanner does not process acknowledgement without response data from target ports and maintain no state at all. When receiving packets from target host after TCP handshakes, ZBanner employs an allowlist to filter out packets we do not care. The following TCP flag combinations can carry response data:

\begin{enumerate}
	\item\textit{ACK}: Indicates a large-size response and this is the first packet with data that containing banner.
	\item\textit{PSH-ACK}: Contains the whole response data to our probe.
	\item\textit{FIN-PSH-ACK}: Includes the whole response data and actively closes the connection.
\end{enumerate}

\subsection{Handling TCP Sliding Window Protocol}

\sloppy{}

Sliding window protocol controls and optimizes packet flow between the sender and receiver in a TCP connection, while ensuring a balanced approach to packet delivery. The protocol requires the receiver to acknowledge receipt of each data packet, and it enables the receiver to use a single or multiple acknowledgments to confirm the delivery of multiple packets.

However, the sliding window protocol allows sending a batch of packets that makes trouble to ZBanner. ZBanner potentially assumes the first received packet with response data is the actual first packet in order. This is because our stateless scanning model cannot identify the order of packets without maintaining states for each connection.

To handle the TCP sliding window protocol, ZBanner sets a proper TCP window size for ACK packet to mention a limited capability of handling data. This makes target host sending no more than one packet before our acknowledgement in most cases.

\subsection{Extensible Probe Module}

Compared to single-packet scan, ZBanner encounters a diverse range of services over TCP from which richer information can be extracted. ZBanner offers an extensible probe module that supports custom generation of probe content and handling of response results through a unified interface. Once ZBanner establishes a connection with target port after confirming its TCP aliveness with a non-zero window size, the specified probe module will generate proper probe data based on the type of target. Then generated content is sent by ZBanner, and the response banners are processed and reported by the our specified probe module.

This approach significantly simplifies the process of extending ZBanner’s capabilities. By creating custom probe modules, various types of information detection over TCP can be achieved. We have implement some typical probes to achieve practical functions including service identification, service version detection and etc.

\section{Evaluation}\label{sec:evaluation}

\sloppy{}


Since ZBanner is the implementation of our stateless scanning model and the first scanner that establishes TCP connections and obtains responses in completely stateless mode, we conduct a series of experiments to characterize its differences from existing stateless scanners, feasibility in fast large-scale network scanning and performance in terms of scan rate and memory usage.
First, we investigated whether ZBanner can handle scan task at high-speed scan rate through the relationship between port hit rate, banner respond rate and scan rate.
Second, we computed the coverage of ZBanner on scan targets to investigate whether it is feasible for common network measurements.
Then, we compared ZBanner with the state-of-the-art solutions in same condition to present its characteristics and advantages.
Last, we showed the real performance of ZBanner and other solutions and in large-scale scanning scenarios


\subsection{Experimental Setup}\label{sec:experimentset}

All experiments were conducted on a virtual cloud host equipped with a 4-core CPU and 8 GB of memory—a configuration commonly used in personal cloud hosting. The host was allocated symmetric 1 GbE bandwidth for both upstream and downstream communication. The operating system running on the host was Ubuntu 22.04 LTS, with the 6.5.0-14-generic version of the Linux kernel. The server’s geographical location is within Chinese mainland, and no other special network configurations were applied.

Since information obtaining over TCP could be various types in different scenarios and it is impossible to evaluate them all, we take the banner grabbing as the most basic and representative example of information obtaining over TCP.

We would like to characterize the efficiency of the various solutions in terms of time by the average scan rate. However, the average scan rate is very difficult to compute accurately, so we use the scan duration to most directly characterize the efficiency.

We consider the following state-of-the-art solutions and their respective settings for comparing:

\begin{enumerate}	
	\item \textbf{ZMap/LZR}:
	Conducts a continuous two-phase scan.
	The first phase employs asynchronous, stateless TCP SYN scan using the system calls of raw socket to identify open ports.
	The second phase performs asynchronous, stateful banner grabbing using the libpcap library.
	In our experiments, we selectively enabled only one Handshake module as probe, given that LZR is primarily used for service recognition.
	We utilized the latest release versions of ZMap 3.0 and LZR.
	
	\item \textbf{ZMap/ZGrab}:
	Executes a seperate two-phase scan.
	Similar to ZMap/LZR, the first phase involves asynchronous, stateless TCP SYN scan to identify open ports.
	The second phase leverages Go language goroutines and the \verb|Net| library for asynchronous, stateful banner grabbing.
	In our experiments, we avoided ZGrab from doing TLS handshakes for fairness.
	We used the most recent release versions of ZMap 3.0 and ZGrab 2.
	
\end{enumerate}

To minimize interference from geographical location and provider restrictions on detection results, all scanning in our experiments focused on a collection of publicly accessible IP addresses managed by Chinese mainland operators, referred to as the “address set”. This address set encompasses 6,691 autonomous systems and approximately 343 million unique public IPs.

\subsection{Scan Rate: Is It Too Fast over TCP?}

In previous studies\cite{zmap}, it has been demonstrated that when performing stateless port scanning, the scan rate (i.e., the rate at packets are sent per second) does not significantly affect the number of port hits (i.e., the count of received SYN-ACK responses) at lower than 1 gbE network speeds. However, when ZBanner establishes TCP connections using a stateless approach and sends probes to target ports, more round-trip exchanges of TCP packets are involved. Consequently, it becomes crucial to determine whether the number of response banner over TCP is influenced by variations in scan rate.

\begin{figure}[htbp]
	\centering
	\includegraphics[width=1.0\columnwidth]{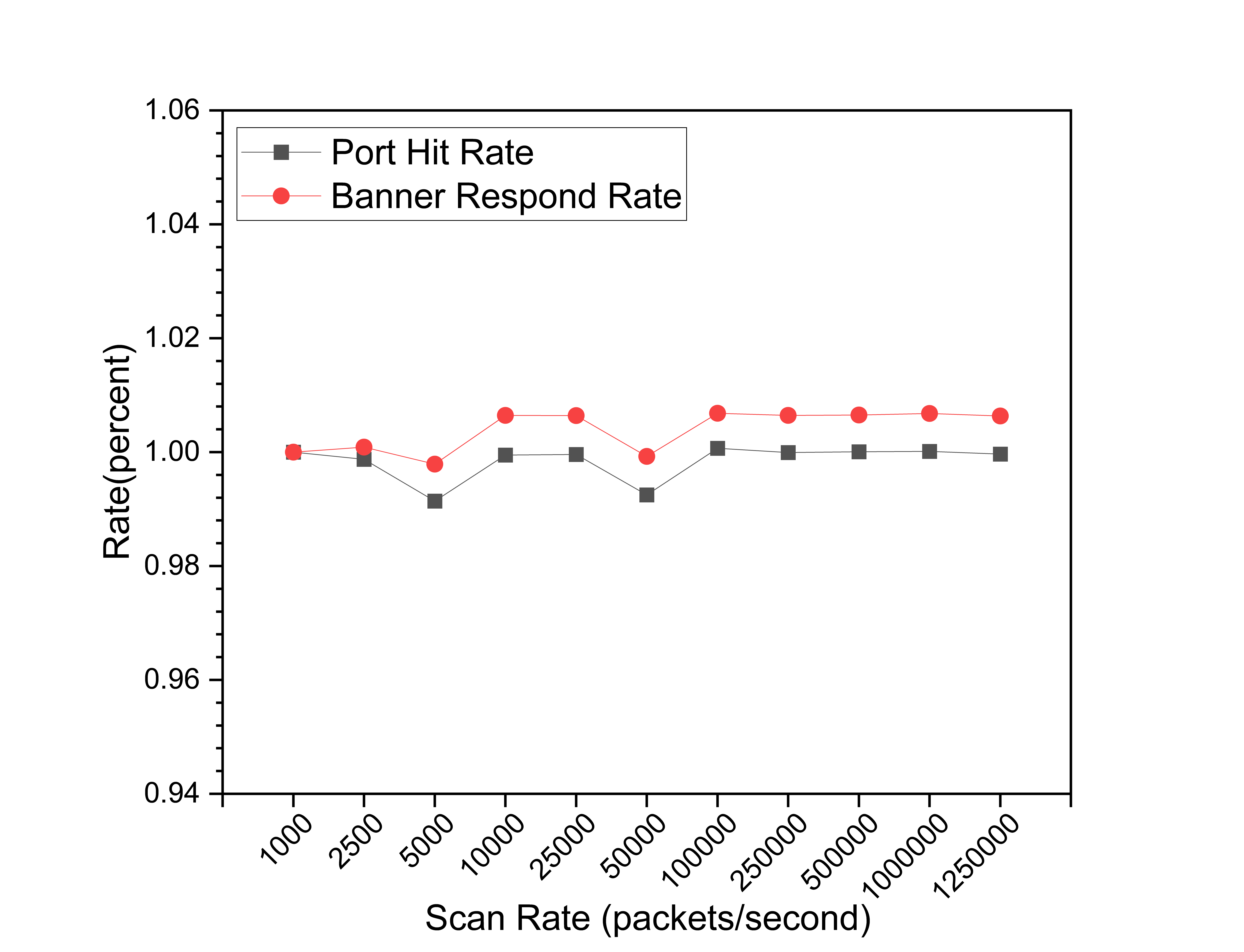}
	\caption{Relationship Between Port Hit Rate, Banner Respond Rate and Scan Rate}
	\label{fig:hitrate}
\end{figure}

To address this, we conducted a series of tests using ZBanner. By limiting the packets sent per second, we attempted to grab banners from port 80 across the address set. We performed 10 trials, varying the scan rate from 1,000 packets per second (pps) to 1.25 million pps. For each scan rate, we calculated the mean port hit count and response banner count. We take the results obtained at slowest scan rate (1,000 pps) as baseline and plot the relationship between port hit rate and banner respond rate as shown in Figure \ref{fig:hitrate}.

Our result reveals a clear positive correlation between port hit rate and banner respond rate. This relationship arises because whether a port is hit successfully directly determines whether the follow-up TCP connection is tried to establish.
Surprisingly, statistical analysis shows no significant correlation between scan rate and either port hit rate or banner respond rate. This suggests that ZBanner remains unaffected by changes in scan rate even working on completely stateless manner.
Consequently, running ZBanner at high-speed scan rates within the bandwidth limit of 1 gbE does not compromise port hit rate or banner respond rate.
Furthermore, our results validate that ZBanner, based on our stateless scanning model, is fully capable of handling responses triggered by gigabit-speed packet transmission.

\subsection{Coverage: How Many Banners We Have Lost?}

While high scan rate does not lead to a decrease in port hit rate and banner respond rate, it fails to fully describe the coverage of information obtaining over TCP during stateless banner grabbing. Specifically, we need to understand what proportion of banners ZBanner can effectively grab.

Given the vast number of internet ports and their dynamic states, determining the exact count of internet ports that respond to a specific probe during a given time frame (i.e., the number of ports running a particular service) is challenging almost impossible.
To address this, we employed ZBanner to try to establish multiple connections in sequence for each target port within a single turn of scanning. Every time the connection established, we sent the same type of probe to attempt to grab banners.
To avoid interference, we ensured a minimum 1-second interval between connection attempts for each target port.
By analyzing the distribution of unique response banners relative to the number of connection attempts, we estimated the actual count of ports running the specified service during that time frame.

As the number of connection attempts increases, the port hit count eventually converges\cite{zmap}. We expect a similar convergence in the response banner count. If convergence occurs, we can treat the converged value as an estimate of the actual number of service-running ports during that period. This estimate serves as a baseline for assessing banner grabbing coverage at lower connection attempt rates.

\begin{figure}[htbp]
	\centering
	\includegraphics[width=1.0\columnwidth]{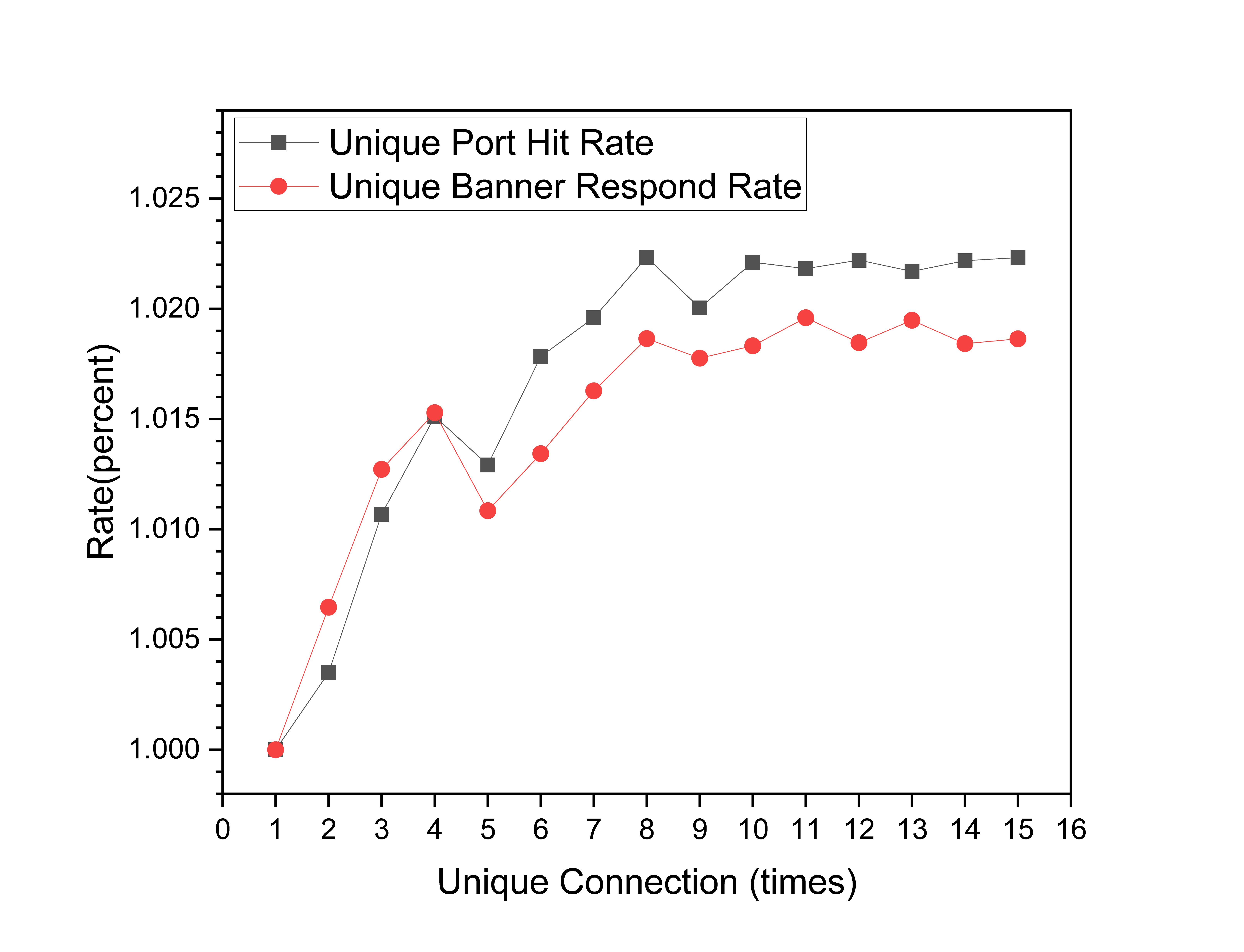}
	\caption{Relationship Between Port Hit Rate, Banner Respond Rate and Connection Times}
	\label{fig:coverage}
\end{figure}

We randomly sampled 30\% of addresses from the address set and conducted 15 rounds of ZBanner scans on port 80 for each sampled address. The connection attempts for each target port ranged from 1 to 15 incrementally.
After each connection, we recorded the unique port hit count and response banner count. Since the final convergence value is unknown, we used the results from the first scan (1 attempt per port) as the baseline. We calculated port hit rates and banner respond rates for subsequent rounds to analyze their trends as shown in Figure \ref{fig:coverage}.

\begin{table}[htbp]
	\centering
	\caption{Coverage for the Initial 3 Rounds}
	\label{tab:coverage}
	\begin{tabular}{ccc}
		\hline
		\multirow{2}{*}{\textbf{Round}} & \multicolumn{2}{c}{\textbf{Coverage}}         \\
		& \textbf{Port Hit} & \textbf{Banner Grabbing} \\ \hline
		1                               & 97.84\%             & 98.15\%                     \\
		2                               & 98.18\%             & 98.79\%                     \\
		3                               & 98.89\%             & 99.40\%                     \\ \hline
	\end{tabular}
\end{table}

Based on the observed trend in banner respond rate, we notice that the results stabilize from the 10th round onward. Consequently, we consider the average count of response banners from rounds 10 to 15 as the convergence value. Using this value, we estimate the coverage for the initial 3 rounds, as shown in the Table \ref{tab:coverage}.
\footnote{The port hit coverage obtained using ZBanner closely aligns with the estimates by Durumeric et al.\cite{zmap}, with minor differences. We attribute these discrepancies to Durumeric’s use of “back-to-back” SYN packets to increase the number of attempts, whereas ZBanner intentionally introduces a minimum 1-second interval between connection attempts to avoid interference. Additionally, transient network unreachability may cause the loss of consecutive data packets. Therefore, theoretically, ZBanner’s scan results provide a more accurate estimate of coverage.}
While the results may exhibit slight deviations due to factors such as sample selection, scan start time, port services, and probe types, empirical evidence from multiple experiments suggests that as long as upstream network bottlenecks are not a concern, the banner grabbing coverage achieved by attempting only one connection remains consistently around 98\%.

This result underscores that ZBanner’s default mode (attempting only one connection) yields sufficiently comprehensive outcomes for common use cases. While it is possible to enhance coverage by configuring ZBanner to increase the number of connection attempts, this approach inevitably extends scan duration.

Unlike port scanning, which requires only a single round-trip packet exchange, obtaining information over TCP involves establishing complete TCP connections and engaging follow-up data interactions. This process entails multiple round-trip exchanges of packets within the network. ZBanner, deliberately forgoes any guarantees of reliable packet transmission while operating under stateless scanning conditions.
Consequently, banner grabbing coverage under stateless conditions should be significantly lower than port hit coverage in theory.
However, experimental results as shown in Figure \ref{fig:coverage} and Table \ref{tab:coverage} reveal an intriguing relationship: while banner respond rate and port hit rate converge due to their interdependence, the former’s convergence value remains lower than the latter’s. This discrepancy challenges our intuitive understanding.

Two key factors contribute to this phenomenon:

\begin{table*}[htbp]
	\centering
	\caption{Mitigation of Packet Loss}
	\label{tab:maintainstate}
	\begin{tabular}{cccl}
		\hline
		\textbf{Type} & \textbf{Direction}                         & \textbf{If Mitigated}       & \multicolumn{1}{c}{\textbf{Reason}}                                                              \\ \hline
		SYN           & ZBanner $\rightarrow$ Target & \faTimes & No maintenance of state.                                                                \\
		SYN-ACK       & ZBanner $\leftarrow$ Target  & \faCheck & Target didn't receive ACK for this packet and retransmit. \\
		ACK with data & ZBanner $\rightarrow$ Target & \faCheck & Target didn't receive this packet and retransmit the previous packet.    \\
		Banner        & ZBanner $\leftarrow$ Target  & \faCheck & Target didn't receive ACK for this packet and retransmit. \\
		RST           & ZBanner $\rightarrow$ Target & \faCheck & Target didn't receive this packet and retransmit the previous packet.    \\ \hline
	\end{tabular}
\end{table*}

\textbf{State Maintenance from one Endpoint}: Although ZBanner does not maintain state to ensure reliable packet transmission, the target hosts actively maintain the TCP connection state once they received the TCP SYN packet. From ZBanner’s perspective, this one-sided effort to maintain network transmission reliability effectively mitigates most packet loss scenarios (as indicated in the Table \ref{tab:maintainstate}). Consequently, banner grabbing coverage does not significantly lag behind port hit coverage.

\textbf{Open Ports Unresponsive to Expected Services}: Izhikevich et al.\cite{lzr} measured that a considerable portion of TCP ports are failed to complete the expected application-layer handshake even after responded SYN-ACK packets. For instance, approximately 14\% of port 80 responses do not conform to HTTP probes. These deviations may arise from non-HTTP protocols running on the port or other anomalies (as listed in Table \ref{tab:exceptions}). While these ports do not impact ZBanner’s scan rate, they do contribute to a lower convergence value in banner respond rates. Consequently, the estimated banner grabbing coverage slightly exceeds port hit coverage.

Furthermore, the purpose of estimating banner grabbing coverage is to evaluate ZBanner’s detection effectiveness and analyze the feasibility of information obtaining over TCP in stateless manner for large-scale network scanning scenarios. For the scope of this study, we focus on scenarios where both the scanning source and samples are located within Chinese mainland. Complex and variable external factors such as scan source location, vendor restrictions, temporary blocks, and transient interruptions, as discussed by Gerry et al.\cite{originscan}, are not considered in our analysis.

\subsection{Comparison: What Does Statelessness Bring to Us?}\label{sec:comparison}

When evaluating the performance of large-scale information obtaining over TCP, several factors need to be considered. Notably, the proportion of open ports in the scanned samples significantly impacts the performance of stateful scanners such as LZR and ZGrab.

To comprehensively assess ZBanner’s performance compared to other state-of-the-art solutions across different scenarios, we followed these steps:

\begin{enumerate}
	\item We randomly selected 500,000 independent addresses from the address set, forming subset A. The remaining addresses constituted the pool for port scanning, from which we identified targets with port 80 open as subset B.
	\item By combining addresses from subsets A and B, we created 11 sample sets. Each set represented a different proportion (ranging from 0\% to 100\%) of known open ports. The total number of unique addresses in each sample set remained constant at 500,000.
	\item These sample sets corresponded to three scenarios:
		\begin{itemize}
			\item Generic Target Set with Unknown Information: Addresses with no prior information about open ports.
			\item Specific Target Set with Known Open Ports: Addresses with confirmed open ports.
			\item Intermediate Target Set with Varying Port Knowledge: Addresses changes after confirmed open ports and is in an intermediate state.
		\end{itemize}
\end{enumerate}

To ensure a fair comparison, we controlled for other factors:
\begin{itemize}
	\item Both ZBanner and ZMap were configured with a single packet-sending thread.
	\item We limited available CPU cores to one using Linux GRUB boot parameters.
\end{itemize}
It has been verified that under these conditions, ZBanner and ZMap achieve comparable maximum packet scan rates for TCP SYN packets sending.

\begin{figure}[htbp]
	\centering
	\includegraphics[width=1.0\columnwidth]{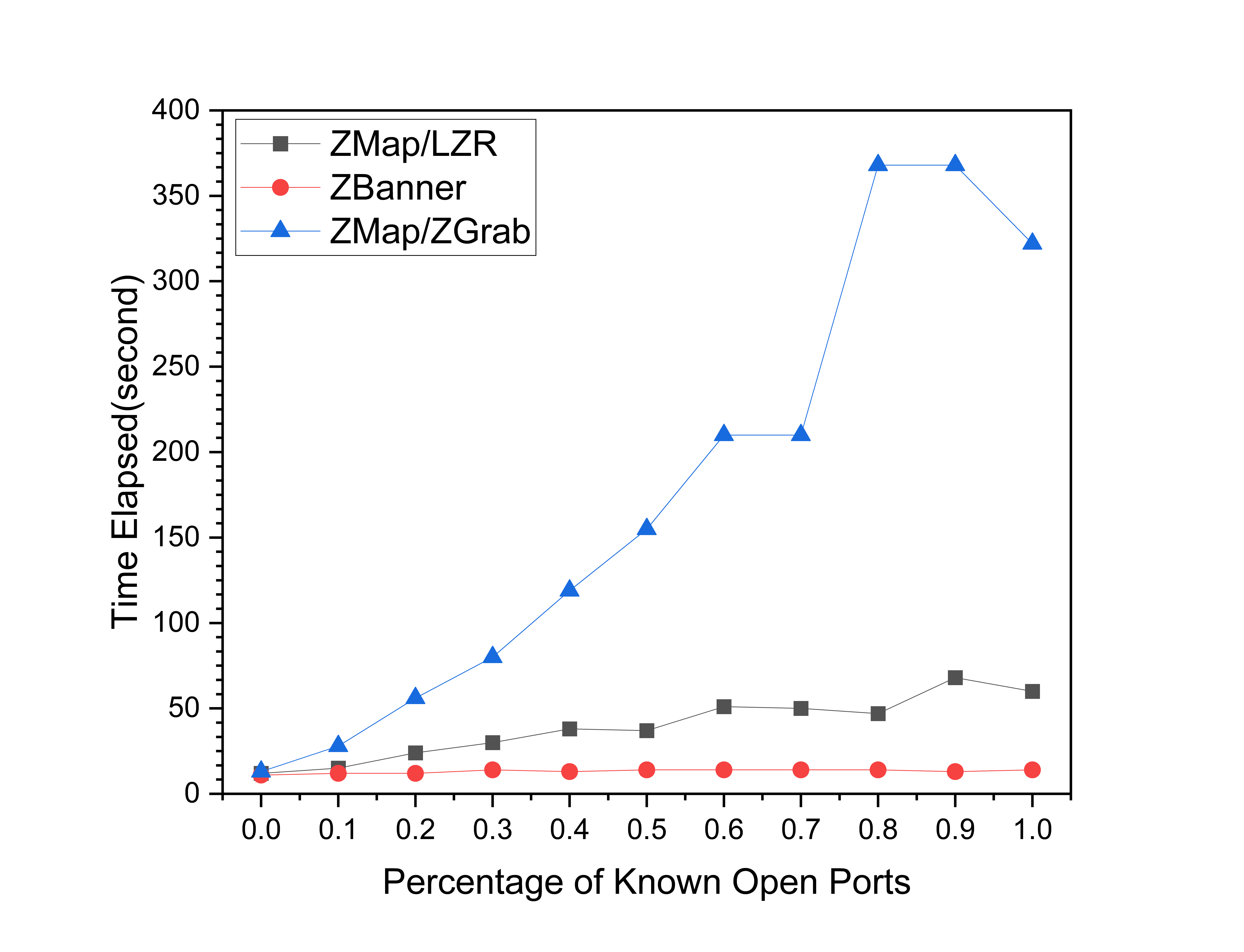}
	\caption{Relationship Between Scan Duration and Known Open Port Proportion}
	\label{fig:comparisontime}
\end{figure}

\begin{figure}[htbp]
	\centering
	\includegraphics[width=1.0\columnwidth]{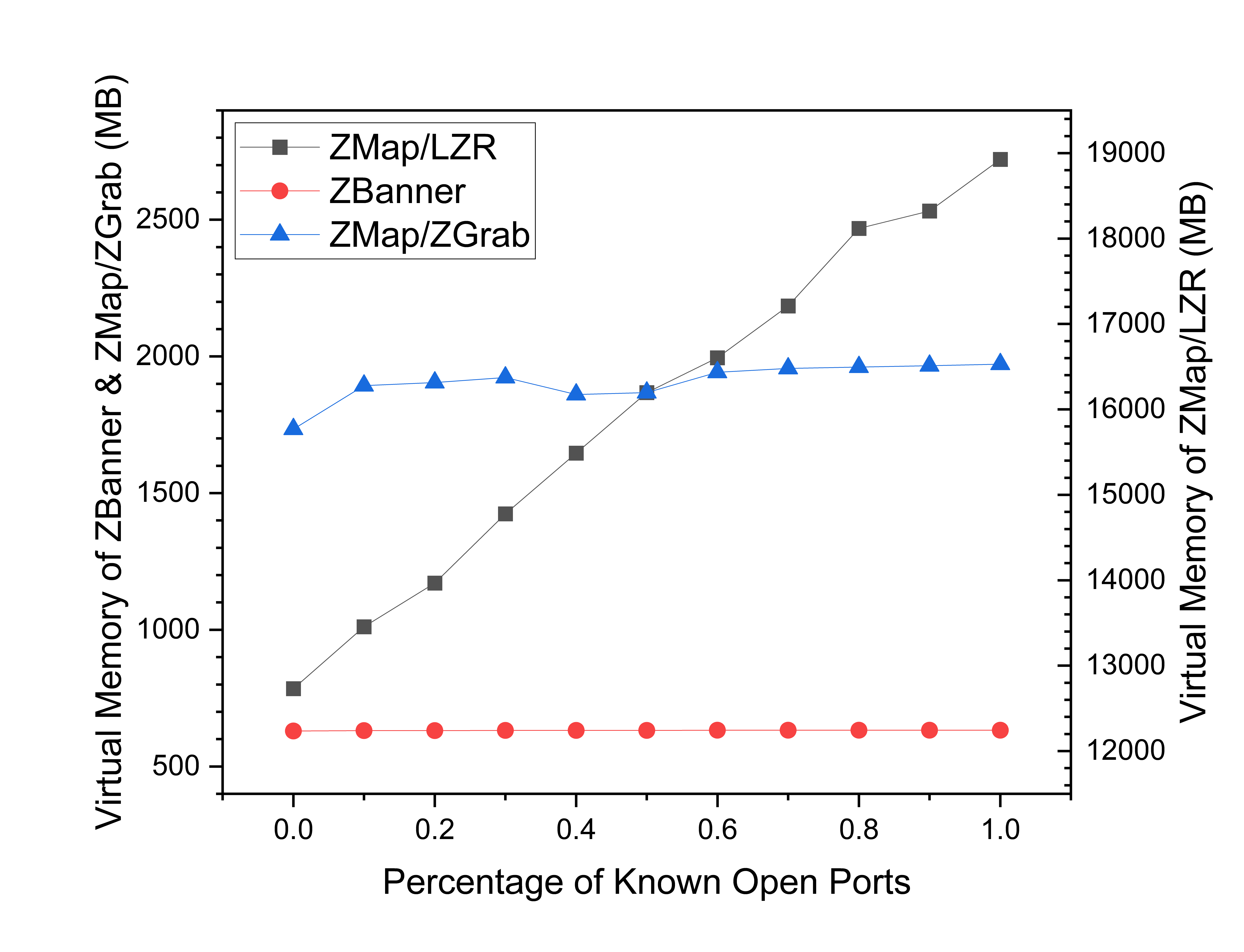}
	\caption{Relationship Between Virtual Memory Usage and Known Open Port Proportion}
	\label{fig:comparisonmemory}
\end{figure}

Under these experimental conditions, we scanned and grabbed banners using ZBanner and other tools. All scans operated at their maximum scan rate (ensuring no packet loss on operation system and completing the scan tasks). The scan durations for each solution are depicted in Figure \ref{fig:comparisontime}. Additionally, we recorded the user-space virtual memory size occupied by each solution (Figure \ref{fig:comparisonmemory}).

\textbf{Scan Duration}:
When the known open port proportion was 0\%, the performance of all three solutions was comparable. 
As the known open port proportion increased, the performance of ZMap/LZR and ZMap/ZGrab declined. ZMap/ZGrab exhibited the most significant degradation.
ZMap/LZR’s scan duration was up to 5.2 times longer than ZBanner, while ZMap/ZGrab’s was 28.3 times longer.
ZBanner consistently maintained high performance even when the known open port proportion reached 100\%. Its stateless packet exchange approach avoided the connection state overhead faced by ZMap/LZR and ZMap/ZGrab.

\textbf{Memory Usage}: ZBanner’s low memory usage aligned with expectations due to our stateless scanning model.
ZMap/LZR’s memory usage increased significantly (up to 18.9 GB) as it maintained connection states in user space.
ZMap/ZGrab’s memory usage remained stable because ZGrab leveraged Go language’s \verb|Net| standard library, allowing the operating system protocol stack to manage connection states and preventing memory growth from impacting user memory space. However, this stability comes at the cost of scan rate limitations imposed by the operating system protocol stack, resulting in significantly increased scan duration.

In summary, ZBanner performs comparably to current state-of-the-art solutions when the known open port proportion is 0\%. As the proportion increases, ZBanner consistently maintains high performance and low memory usage, outperforming other tools that suffer from performance degradation and increased memory consumption.

\subsection{Practice: Real Performance in Large-Scale Scanning Scenarios}

\begin{table*}[t]
	\centering
	\caption{Comparison of Performance in Large-Scale Scanning Scenarios}
	\label{tab:comparisonPractice}
	\begin{tabular}{c|c|c|c}
		\hline
		\textbf{Type of Targets}                                    & \textbf{Solution} & \textbf{Scan Duration} & \textbf{Virtual Memory Usage} \\ \hline
		\multirow{3}{*}{Generic Targets with Unknown Open Ports}    & \textbf{ZBanner}  & \textbf{5m32s}         & \textbf{447MB}                \\ \cline{2-4} 
		& ZMap/LZR          & 29m11s                 & 223MB+23.2GB                  \\ \cline{2-4} 
		& ZMap/ZGrab        & 21m3s                  & 223MB+2209MB                  \\ \hline
		\multirow{3}{*}{Specific Targets with Confirmed Open Ports} & \textbf{ZBanner}  & \textbf{19s}           & \textbf{428MB}                \\ \cline{2-4} 
		& ZMap/LZR          & 31m                    & 996MB+22.7GB                  \\ \cline{2-4} 
		& ZMap/ZGrab        & 29m50s                 & 996MB+2215MB                  \\ \hline
	\end{tabular}
\end{table*}

Building upon the work of Izhikevich et al.\cite{lzr}, it has been observed that when ZMap/LZR operates in an environment with 1 GbE bandwidth and optimal hardware conditions, its scan rate surpasses that of ZMap in identifying live hosts on the generic ports across the internet. This difference is particularly pronounced for low-hit-rate ports. This observation aligns with the experimental findings in Section \ref{sec:comparison}, where ZBanner’s advantage from stateless scanning model is not fully evident when the proportion of known open ports is zero.

However, it is essential to recognize that ideal hardware conditions are not always achievable in large-scale network measurements. To demonstrate ZBanner’s practical performance under commonly used hardware conditions, we conducted scans using ZBanner and the current state-of-the-art solutions on the whole address set targeting port 80. We then curated a subset C comprising approximately 3.61 million unique addresses with known open ports. Both solutions were configured to operate at their maximum scan rates, ensuring complete execution without packet loss on operation system.
The scan duration and virtual memory usage for both solutions during these two rounds of scanning are summarized in Table \ref{tab:comparisonPractice}.

When scanning a generic port set of internet targets, ZBanner’s scan rate is approximately 5.2 times faster than ZMap/LZR and 3.8 times faster than ZMap/ZGrab according to their scan duration. For specific targets with confirmed open ports, ZBanner’s scan rate is more than 90 times faster than both ZMap/LZR and ZMap/ZGrab.

ZBanner maintains consistently low memory usage. In contrast, ZMap/LZR’s memory consumption exceeds 20 GB when scanning a diverse set of internet targets. However, this memory usage does not increase proportionally with the known open port ratio. This limitation arises from the hardware configuration used in our experiments, where memory usage is already at system limit. Attempting to increase scan rates further would result in excessive resource utilization by the process, leading to termination by the operating system.

In summary, when conducting large-scale information obtaining over TCP under commonly used hardware conditions, ZBanner exhibits significant advantages over the current state-of-the-art solutions in terms of both scan rate and memory usage.

\section{Discussion}
\sloppy{}

We have verified the feasibility and demonstrated the characteristics of ZBanner through experiments. However, due to the variety of scanning tasks in real world, the advantages of ZBanner in practical applications go beyond that. At the same time, ZBanner also has some limitations due to its operation mechanism from the stateless scanning model.
In this section, we discuss the advantages and limitations of the stateless scanning model and ZBanner.

\subsection{Advantages in Practice}

ZBanner is the implementation of our stateless scanning model and also a modular stateless scanner for large-scale network measurement. From the evaluation results, ZBanner has similar properties to previous stateless scanners, but its scanning efficiency in terms of time is affected by the percentage of open ports in the target address set: the higher the percentage of open ports, the greater the advantage of ZBanner over similar solutions. But at worst it can fall back to the same level as existing stateless port scanners. In addition, ZBanner provides a one-stop solution for various two-phase scans.

Although the efficiency gains of ZBanner may seem modest in absolute terms based on our experiments, the significance of ZBanner’s efficiency advantage is magnified in practical measurements.
This is not only because ZBanner provides a one-stop solution for port scanning and information obtaining over TCP,  but also the specific task may require establishing TCP connections and sending probes for multiple times to one target port.
For example, we have implemented an LZR like probe to identify service and it takes from 1 to 5 times of connections for every target on optimal handshake order.

Further, for some scanning tasks with multiple processes, the efficiency of ZBanner increases more.
Taking the active TLS server fingerprinting as an example, we implemented the corresponding probe to extract the most classical JARM fingerprint\cite{jarm}. To extract a JARM fingerprint, 10 specific probes are sent to the target port and obtain corresponding responses.
We believe that the following steps are actually included in the process of fingerprint extraction:
\begin{enumerate}
	\item Confirm whether the target port is open or not.
	\item If the target port is open, send a probe to confirm that it is running TLS services.
	\item If the target port is running a TLS service, connect to it multiple times to send probes and obtain responses.
\end{enumerate}
ZBanner is able to combine all the steps in one round of scanning unlike all existing tools of its kind. Since the execution of any step depends on the result of the previous step, bandwidth and other resources are not wasted. Existing stateful scanners and two-phase scan solutions are not well suited to accomplish this task for large-scale targets. For ZBanner, all we need is to write a probe.

\subsection{Inherent Limitations}

Previous internet network measurements relies on stateless scanners to do port scanning as the first step of surveys, because only stateless scanners could handle such large-scale targets in an acceptable time.
But the results from port scanning contain less information and can be misleading.
Therefore, slower stateful scanners must be introduced to the second phase scanning for richer information.
Our stateless scanning model allows obtaining responses over TCP in a high-speed scan rate.
However, ZBanner is intended to enhance the capabilities of existing stateless scanners, rather than completely replace some stateful scanners. This is because the inherent limitations from the stateless scanning model.

First, since the stateless scanning model can only recognize the first packet with payload returned by the target host, the banner we obtain in a given case is incomplete.
A typical example is that the response banner we obtained is sufficient to recognize the HTTP protocol and identify its version, server, cookie, etc.
But sometimes it fails to obtain the complete HTTP header for further detection.
And most of the time, it's impossible to get whole web pages.
The good thing is that our model is sufficient for identifying most protocols and various scanning tasks.
Scanning and detection at the content level is beyond the capabilities of ZBanner, and we believe it is better accomplished with existing stateful scanners.

Second, since the stateless scanning model can only complete one-time data interaction after completing the TCP handshake, for the TLS protocol, ZBanner can only identify and obtain some information about the TLS itself, but cannot touch the upper layer services.
While at some point it is possible to identify upper layer services by the ALPN\cite{alpn_rfc7301} field of the TLS, we recommend the use of specialized tools for further detection.
\\

Anyhow, stateful scanners maintain complicate connection states but could finish more data exchange.
Our stateless scanning model is like a balance between existing stateless scanners and stateful scanners.
The advent of the stateless scanning model opens up more possibilities for large-scale network measurements, and there are many tasks that can be optimized with a tool like ZBanner.
We leave this as future work.

\section{Conclusion}

\sloppy{}

This paper presents a novel analysis of existing stateless scanning techniques from the perspective of TCP finite-state machines(FSM). Leveraging a simplified TCP FSM, we construct a stateless scanning model which are capable of establishing complete TCP connections and obtaining further information over TCP. Building upon this model, we design and implement ZBanner, a modular stateless scanner specifically tailored for large-scale network measurements. ZBanner operates entirely in a stateless manner, significantly enhancing the efficiency of information obtaining over TCP in large-scale networks. However, due to the limited interaction over TCP, it remains challenging to detect services running on the TLS encrypted protocol. Future work will focus on efficiently scanning services over the TLS encrypted protocol with the stateless scanning model.

%
%
%


\bibliographystyle{ACM-Reference-Format}
\bibliography{zbanner_ref_data}

\appendix

\section{Ethics}

\sloppy{}

This work does not raise any ethical issues. We do everything in our power to avoid causing problems for network operators, users, and local network administrators while large-scale scanning. We confirmed with cloud network provider had sufficient capacity for our high-speed experiments and randomize the target addresses to avoid overwhelming destination networks. We accept all complaints during the scanning and create filter lists.

In fact, for most application services, our scans are closer to real user access because we create normal TCP connections and close them in a timely manner, rather than leaving some half-connected state as existing stateless scanners may do.


\section{Numerical Data of Experiments}

\sloppy{}

These are numerical data of experiments mentioned in Section \ref{sec:evaluation}. We cannot provide IP addresses or ports for privacy of network operators and users.

The relationship between number of port hit, number of response banner 

\begin{table}[htbp]
	\centering
	\caption{Relationship Between Port Hit, Banner Respond and Scan Rate}
	\label{tab:hitrate}
	\begin{tabular}{c|c|c}
		\hline
		\textbf{Scan Rate (p/s)} & \textbf{Port Hit} & \textbf{Banner Respond} \\ \hline
		1000                     & 3528941           & 3228723                 \\ \hline
		2500                     & 3524516           & 3231536                 \\ \hline
		5000                     & 3498664           & 3221968                 \\ \hline
		10000                    & 3527165           & 3249572                 \\ \hline
		25000                    & 3527474           & 3249465                 \\ \hline
		50000                    & 3502488           & 3226279                 \\ \hline
		100000                   & 3531393           & 3250744                 \\ \hline
		250000                   & 3528736           & 3249572                 \\ \hline
		500000                   & 3529231           & 3249771                 \\ \hline
		1000000                  & 3529403           & 3250608                 \\ \hline
		1250000                  & 3527735           & 3249222                 \\ \hline
	\end{tabular}
\end{table}

\begin{table}[htbp]
	\centering
	\caption{Relationship Between Port Hit, Banner Respond and Connection Times}
	\label{tab:coverageDetail}
	\begin{tabular}{c|c|c}
		\hline
		\textbf{Connect Times} & \textbf{Port Hit} & \textbf{Banner Respond} \\ \hline
		1                      & 449174            & 384699                  \\ \hline
		2                      & 450747            & 387186                  \\ \hline
		3                      & 453973            & 389593                  \\ \hline
		4                      & 455966            & 390581                  \\ \hline
		5                      & 454976            & 388870                  \\ \hline
		6                      & 457190            & 389867                  \\ \hline
		7                      & 457976            & 390963                  \\ \hline
		8                      & 459209            & 391874                  \\ \hline
		9                      & 458179            & 391531                  \\ \hline
		10                     & 459108            & 391750                  \\ \hline
		11                     & 458977            & 392238                  \\ \hline
		12                     & 459152            & 391802                  \\ \hline
		13                     & 458920            & 392195                  \\ \hline
		14                     & 459137            & 391784                  \\ \hline
		15                     & 459202            & 391869                  \\ \hline
	\end{tabular}
\end{table}

\begin{table*}[htbp]
	\centering
	\caption{Comparison of Scan Duration and Memory Usage}
	\label{tab:compare}
	\begin{tabular}{c|c|c|c|c}
		\hline
		\textbf{Percentage of Known Open Ports} & \textbf{Solution} & \textbf{Time Elapsed(second)} & \textbf{Virtual Memory} & \textbf{Resident Memory} \\ \hline
		\multirow{3}{*}{0\%}                      & ZBanner           & 11                            & 630M                    & 42M                      \\ \cline{2-5} 
		& ZMap/LZR          & 12                            & 231M+12.5G              & 6M+211M                  \\ \cline{2-5} 
		& ZMap/ZGrab        & 13                            & 231M+1503M              & 6M+34K                   \\ \hline
		\multirow{3}{*}{10\%}                     & Zbanner           & 12                            & 631M                    & 50M                      \\ \cline{2-5} 
		& ZMap/LZR          & 15                            & 255M+13.2G              & 38M+949M                 \\ \cline{2-5} 
		& ZMap/ZGrab        & 32                            & 255M+1638M              & 38M+162M                 \\ \hline
		\multirow{3}{*}{20\%}                     & Zbanner           & 12                            & 631M                    & 50M                      \\ \cline{2-5} 
		& ZMap/LZR          & 24                            & 268M+13.7G              & 53M+1556M                \\ \cline{2-5} 
		& ZMap/ZGrab        & 57                            & 268M+1637M              & 53M+150M                 \\ \hline
		\multirow{3}{*}{30\%}                     & Zbanner           & 14                            & 632M                    & 50M                      \\ \cline{2-5} 
		& ZMap/LZR          & 30                            & 279M+14.5G              & 65M+2187M                \\ \cline{2-5} 
		& ZMap/ZGrab        & 80                            & 279M+1644M              & 65M+148M                 \\ \hline
		\multirow{3}{*}{40\%}                     & Zbanner           & 13                            & 632M                    & 50M                      \\ \cline{2-5} 
		& ZMap/LZR          & 38                            & 289M+15.2G              & 75M+2869M                \\ \cline{2-5} 
		& ZMap/ZGrab        & 119                           & 289M+1572M              & 75M+131M                 \\ \hline
		\multirow{3}{*}{50\%}                     & Zbanner           & 14                            & 632M                    & 51M                      \\ \cline{2-5} 
		& ZMap/LZR          & 37                            & 297M+15.9G              & 83M+3430M                \\ \cline{2-5} 
		& ZMap/ZGrab        & 155                           & 297M+1571M              & 83M+127M                 \\ \hline
		\multirow{3}{*}{60\%}                     & Zbanner           & 14                            & 633M                    & 51M                      \\ \cline{2-5} 
		& ZMap/LZR          & 51                            & 305M+16.3G              & 91M+4245M                \\ \cline{2-5} 
		& ZMap/ZGrab        & 210                           & 305M+1637M              & 91M+126M                 \\ \hline
		\multirow{3}{*}{70\%}                     & Zbanner           & 14                            & 633M                    & 51M                      \\ \cline{2-5} 
		& ZMap/LZR          & 50                            & 312M+16.9G              & 98M+4584M                \\ \cline{2-5} 
		& ZMap/ZGrab        & 210                           & 312M+1644M              & 98M+127M                 \\ \hline
		\multirow{3}{*}{80\%}                     & Zbanner           & 14                            & 633M                    & 51M                      \\ \cline{2-5} 
		& ZMap/LZR          & 47                            & 318M+17.8G              & 102M+5803M               \\ \cline{2-5} 
		& ZMap/ZGrab        & 368                           & 318M+1643M              & 102M+119M                \\ \hline
		\multirow{3}{*}{90\%}                     & Zbanner           & 13                            & 633M                    & 51M                      \\ \cline{2-5} 
		& ZMap/LZR          & 68                            & 323M+18G                & 102M+5230M               \\ \cline{2-5} 
		& ZMap/ZGrab        & 368                           & 323M+1643M              & 102M+117M                \\ \hline
		\multirow{3}{*}{100\%}                    & Zbanner           & 14                            & 633M                    & 51M                      \\ \cline{2-5} 
		& ZMap/LZR          & 60                            & 328M+18.6G              & 113M+6945M               \\ \cline{2-5} 
		& ZMap/ZGrab        & 332                           & 328M+1644M              & 112M+128M                \\ \hline
	\end{tabular}
\end{table*}

\section{Process of Simplifying TCP FSM}\label{sec:fsmSimpify}

\sloppy{}

\begin{figure*}[htbp]
	\centering
	\includegraphics[width=2.0\columnwidth]{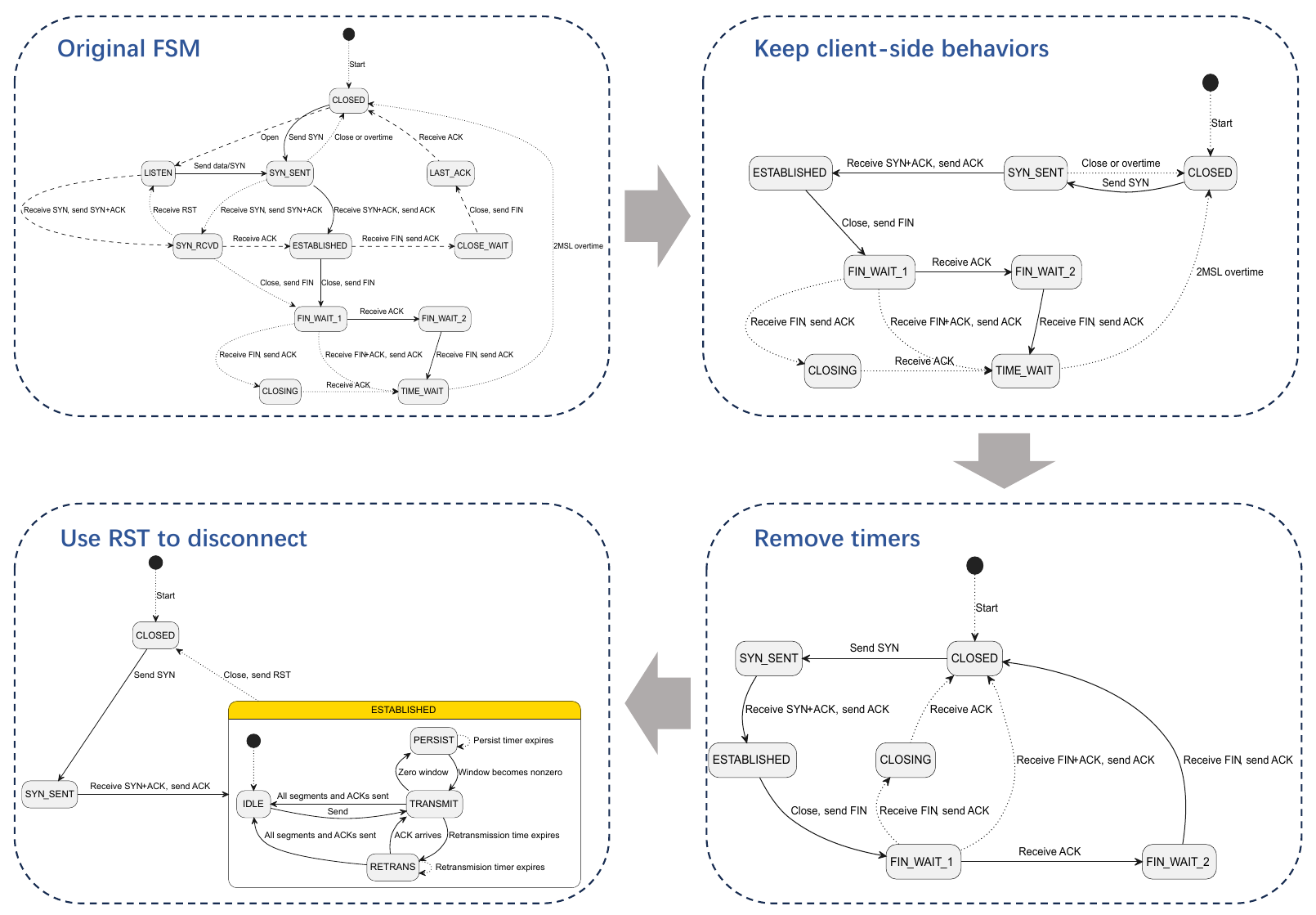}
	\caption{Process of Simplifying FSM}
	\label{fig:fsmSimplyProcess}
\end{figure*}

Following the principles below, our simplification process to TCP FSM is shown in Figure \ref{fig:fsmSimplyProcess} in detailes.

\begin{enumerate}
	\item Since scanning is a client-side behavior, we remove typical server-side state transitions, such as from \verb|LISTEN| to \verb|SYN_RCVD|.
	\item Stateless conditions prevent us from maintaining timers for connections. Consequently, we eliminate state transitions caused by timeouts, such as from \verb|TIME_WAIT| to \verb|CLOSED|.
	\item In the absence of maintaining half-closed connections, we always choose to terminate communication by having the client send an RST packet. This simplifies the process such as from \verb|ESTABLISHED| to \verb|FIN_WAIT_1|.
\end{enumerate}

\end{document}